\documentclass[twocolumn,showpacs,preprintnumbers,amsmath,amssymb]{revtex4}
\usepackage{graphicx}
\usepackage{bm}

\def\ave#1{\langle#1\rangle}

\begin{document}

\title{Rapidly rotating Bose-Einstein condensates in an anharmonic confinement}

\author{Li-Hua Lu and You-Quan Li}
\affiliation{Zhejiang Institute of Modern Physics, Zhejiang
University, Hangzhou 310027, P. R. China}

\begin{abstract}
We examine a rapidly rotating Bose-Einstein condensate in an
anharmonic confinement and find that many properties such as the
critical rotating frequency and phase diagram are quite different
from those in a harmonic trap. We investigate the phase
transitions by means of average-vortex-approximation. We find that
the vortex lattice consists of a vortex array with a hole in the
center of the cloud as the rotating frequency $\Omega$ increases
and the vortex becomes invisible when $\Omega$ reaches some value.
\end{abstract}

 \received{\today} \maketitle

\paragraph{Introduction:}

There have been many studies on Bose-Einstein Condensate (BEC)
rotating with high angular momentum, of which many features
resemble some of the novel solid-state systems such as type-II
superconductor and quantum Hall liquid. In a number of earlier
experiments~\cite{Hodby,Haljan,Madison,Abo-Shaeer} where the gas
is confined in a harmonic trap, the energetically favored state
takes a triangular lattice of singly quantized vortices~\cite{Ho}.
In the case of harmonic trapping potential, the centrifugal force
would have exactly cancelled the trapping force and the system
would have collapsed if $\Omega$ reached to the trap frequency
$\omega_{xy}$. Thus the centrifugal force prevents the rotating
frequency $\Omega$ from being enhanced over $\omega_{xy}$. This
limit makes it impossible for angular momentum to reach the higher
magnitude as expected. To resolve this problem, some theoretical
studies suggest that $\Omega$ is no longer bounded by
$\omega_{xy}$ once an anharmonic term is introduced into the
trapping potential. Vortices are successfully created in
experiment~\cite{Bretin} by adding a quartic term in the trapping
potential and their phase transitions are investigate by varying
the $\Omega$. The experimental images make the theoretical
suggestion that the picture should become richer in an anharmonic
trap more conceivable
~\cite{Kasamatsu,Lundh,Fetter,Fischer,Collin,Simula,Aftalion}.
Indeed, for $\Omega >\Omega_c$ ($\Omega_c$ is the critical
frequency), the gas can only form an array of regular vortices in
a harmonic trap. However, in an anharmonic trap, it can also be a
state of multiple quantization and a mixed state except for the
above state, The mixed state consists of a multiply quantized
vortex (hole) at the center of the cloud and singly-quantized
vortices around it.

Currently, people's interest extends to the fast rotating BEC
which exhibits richer picture. We study the fast rotating BEC in
quadratic-plus-quartic trapping potential in this paper. First we
determine the critical frequency $\Omega_c$ by means of the
variational method which differs from that in a harmonic trap. We
prove that the system in an anharmonic trap can form
multiply-quantized state indeed ~\cite{Jackson}. Then, we extend
the approach proposed by Ho~\cite{Ho} to investigate the phase
transitions between the mixed state and the state with an regular
vortex lattices. Comparing the energies for different $l$, we
obtain the $\Omega_h$ over which a hole is generated at the center
of the cloud. The increase of $\Omega$ will make the hole to
absorb the singly-quantized vortices around it gradually and form
a giant vortex state finally. Because the velocity that the hole
absorbs the singly-quantized vortices becomes larger and larger,
the fragility of the vortex lattice increases. However, recent
theoretical study~\cite{A. Aftalion} found that much longer time
were required for $\Omega\simeq\omega_{xy}$ to reach a
well-ordered lattice. Based on the this argument it is reasonable
that vortex can not be detected for $\Omega\simeq\omega_{xy}$ in
experiment~\cite{Bretin}.


We consider a Bose-Einstein Condensate confined in an anharmonic
trap whose potential is well approximated by a superposition of a
quadratic and a quartic potential: ~\cite{A. D. Jackson}
\begin{equation}
v(\vec{r})\simeq\frac{1}{2}M\omega_z^2z^2+\frac{1}{2}M\omega_{xy}^2r^2+\frac{M\omega_{xy}^2}{2a_\perp^2}\lambda{r^4},
   \quad(\lambda>0),
\end{equation}
where $M$ is the atomic mass; $\omega_z$ and $\omega_{xy}$ are the
frequencies of the harmonic potentials along the $z$-axis and in
the $xy$ plane $\mathbf{r}\equiv (x, y)$ respectively;
$a_\perp=[\hbar/(M\omega_{xy})]^\frac{1}{2}$ denotes the
oscillator length and $\lambda$ a small dimensionless parameter
characterizing the strength of the quartic potential which is
$\lambda\approx10^{-3}$ in the experiment ~\cite{Bretin}. The
energy functional is given by
\begin{equation}\label{eq:energy}
  E[\Psi]=\int{d^3r}\Psi^* (h_z+h_\perp)\Psi
  +\frac{g}{2}\int{d^3r}{|\Psi|}^4,
\end{equation}
where $\Psi$ is the condensate wave function, $h_z$ and $h_\bot$
refer to the single-particle Hamiltonian along $z$-axis and in the
$xy$ plane, namely,
$h_z=-({\hbar^2}/{2M})\nabla_z^2+(M\omega_z^2z^2)/2$ and
$h_\bot=-({\hbar^2}/{2M})\nabla_\bot^2+(M\omega_{xy}^2r^2)/2
 +(M\omega_{xy}^2\lambda{r^4})/({2a_\bot^2})$.

To find the optical form of $\Psi$ for a rapidly rotating system,
we ought to minimize the energy Eq.(\ref{eq:energy}) subject to
two constraints. Since both the total number of particles and the
total angular momentum are constants, two Lagrange multipliers are
introduced such that $\delta (E-\mu N -\Omega l_z )=0$. Here the
chemical potential $\mu$ and the rotation frequency $\Omega$ of
the trap can guarantee the particle number as well as the angular
momentum to be constants ~\cite{Pethick}. We therefore solve the
condensate wave function by minimizing the following
Gross-Pitaevskii functional:
\begin{equation}
K=\int{d^3r}\Psi^\ast[h_z+h_\perp-\Omega{L_z}-\mu]\Psi
+\frac{g}{2}\int{d^3r}{|\Psi|}^4,
\end{equation}
in which $L_z$ is the angular momentum along the positive
direction of $z$-axis ; $g=(4\pi\hbar^2a_{sc})/{M}$  is the
strength of the effective two-body interaction with $a_{sc}$ being
the s-wave scatting length.

\paragraph{Critical rotational frequency- $\Omega_c$:}

In reality, for a large condensate, the density profile along z
satisfies the Thomas-Fermi approximation, which is similar to the
stationary case. Since the gas rotates around the $z$-axis and its
feature in the $xy$ plane is more interesting, we simply assume
that the particle density along z is a constant $\rho$ ~\cite{A.
D. Jackson} so that we have a two dimensional quantum system
(e.g., $z=0$ plane without loss of generality).

It is instructive to consider the limit case $g=0$ and $\lambda=0$
when the hamiltonian in $xy$ plane is similar to that in the
quantum Hall regime. In this case the eigenenergy gives rise to
the Landau Levels for which the wave function $\Phi_m$ for the
lowest Landau Level (LLL) reads:
$\Phi_m\propto{r^m\exp({im\phi})\exp({r^2/2a_\perp^2}})$
~\cite{Subrahmanyam}. In experiment, the $g$ and $\lambda$ are
nonzero but their magnitudes are relatively small, thus we can
solve the problem by means of the variational method. Let us write
the variational wave function as
$\Phi=Cr^m\exp({im\phi})\exp({\beta^2r^2/2a_\bot^2})$ with
variational parameter $\beta$. Here
$C=\beta^{m+1}/[{a_\bot^m}(\pi{a_\bot}^2m!)^{1/2}] $ is the
normalization constant. The energy for state
$\Psi=\sqrt{\rho}\Phi$ is given by (in units of
$\rho\hbar\omega_{xy}$)
\begin{eqnarray}\label{eq:theK}
 K&=&\bigl(\frac{1}{2}(\beta^2+\frac{1}{\beta^2})-\frac{\Omega}{{\omega_{xy}}}\bigr)
  \Gamma_1(m)+\frac{\lambda}{2\beta^4}\Gamma_2(m)
  \nonumber \\
  &&\hspace{2mm}
  +\rho{a_{sc}}\beta^2\Gamma_3(m)+(\frac{\Omega}{{\omega_{xy}}}-\mu),
\end{eqnarray}
where $\Gamma_1(m)={\beta^2}\ave{r^2}/{a_\bot^2}=m+1$ refers to
the rms width of the wave function in units of
${\beta^2}/{a_\bot^2}$,
$\Gamma_2(m)={\beta^4}\ave{r^4}/{a_\bot^4}=(m+1)(m+2)$ the
expectation value of $r^4$, and $\Gamma_3(m)=2\pi a_\bot^2
/\beta^2 \int{d^2 r}\mid\Phi\mid^4 =2m!/(2^{2m}(m!)^2)$ the
interaction contribution. Minimizing the energy with respect to
the variational parameter $\beta$, we obtain the following two
equations with respect to $\beta$ for the cases of $m=0$
(non-vortex configuration) and $m=1$ (one vortex of unit
strength), respectively,
\begin{eqnarray}
   \beta_0^6-\beta_0^2-4\lambda+2\rho{a_{sc}}\beta_0^6&=&0,
   \nonumber
\\ 2\beta_1^6-2\beta_1^2-12\lambda+\rho{a_{sc}}\beta_1^6&=&0.
\end{eqnarray}
Their corresponding energies are
\begin{eqnarray}
K_0&=&\big(\frac{1}{2}(\beta_0^2+\frac{1}{\beta_0^2})-\frac{\Omega}{{\omega_{xy}}}\bigr)
 +\frac{\lambda}{2\beta_0^4}\times2+\rho{a_{sc}}\beta_0^2
      \nonumber \\
   &&\hspace{2mm} +(\frac{\Omega}{{\omega_{xy}}}-\mu),
      \nonumber \\
K_1&=&\big(\frac{1}{2}(\beta_1^2+\frac{1}{\beta_1^2})-\frac{\Omega}{{\omega_{xy}}}\bigr)
   \times2+\frac{\lambda}{2\beta_1^4}\times6+\rho{a_{sc}}\beta_1^2\times\frac{1}{2}
\nonumber \\&&{}+(\frac{\Omega}{{\omega_{xy}}}-\mu).
\end{eqnarray}

The experiment data $N=3\times{10^5}$, $\omega_{xy}=2\pi\times65.6
\textrm{Hz}$, $\omega_z=2\pi\times11.0\textrm{Hz}$, and
$a_{sc}=53$\AA, give rise to $\rho{a_{sc}}\approx20$ ~\cite{A. D.
Jackson}. Thus we have approximately $\beta_0=0.397664$ and
$\beta_1=0.551786$. For $K_0$=$K_1$, we obtain
${\Omega_c}/{\omega_{xy}}=0.2223$ with $\Omega_c$ the critical
frequency. The state without vortex is stable when $\Omega <
\Omega_c$, but when $\Omega > \Omega_c $ the state with vortex is
stable. For $\lambda=0$, the critical frequency is
$\Omega_c/\omega_{xy}=0.136$ which is smaller than that for
$\lambda\neq0$. It has been shown ~\cite{Pethick} that the gas
without interaction in a harmonic trap does not  generate vortex.
The contributions of both interaction and rotating frequency
$\Omega$ result in vortex in the condensate. The larger the value
$\lambda$ is, the smaller the contribution of the interaction term
in the total energy will be. Thus, larger rotating frequency
$\Omega$ is required for generating vortex. We examine further
state configurations with larger units of vortex, viz., for
$m=2,3\cdots$ ~\cite{Subrahmanyam}. The energy $K_2$ is minimized
when $\beta_0=0.642021$. $K_1=K_2$ gives
${\Omega_2}/{\omega_{xy}}=0.7128$. Clearly, when
${{\Omega_2}/{\omega_{xy}}}>0.7128$ the $K_1$ becomes larger than
$K_2$, which implies that the state with doubly quantized vortex
is more stable than the singly quantized vortex. Unlike the  case
in harmonic trap where the state with singly quantized vortex
(unit quantum number) is merely most favorable
~\cite{Subrahmanyam} when $\Omega
> \Omega_c$, the stable state in a anharmonic trap favors vortex with
large quantum numbers (angular momentum). The faster (the larger
$\Omega$) the anharmonic trap rotates, the higher the quantum
number of the vortex will be.

\paragraph{Vortex lattice:}
We consider the state with a large angular momentum by looking at
configurations with a large number of vortices of unit strength
centered with a giant vortex of larger strength. Since the
dynamics along $z$-axis is identical to the stationary case,  the
kinetic energy $\mid\nabla_z\Psi\mid^2$ in $z$ direction is
neglectable in Thomas-Fermi approximation (TFA). Let us write
$K=\int{dz}K(z)$,
\begin{eqnarray}\label{eq:Kz}
K(z)&=&\int{d^2r}\Psi^*\Bigl[-\frac{\hbar^2}{2m}\nabla_\bot^2+\frac{1}{2}
+\frac{M\omega_{xy}^2}{2a_\bot^2}\lambda{r^4}-\Omega{L_z}\nonumber
\\&&{}-\mu(z)\Bigr]\Psi +\frac{1}{2}f\int{d^2r}\mid\Psi\mid^4,
\end{eqnarray}
where $\mu(z)=\mu-{M\omega_z^2z^2}/{2}$. We are now in the
position to calculate $K(z)$. The condensate wave function is
expressed as $\Phi=\sum_{m=0}^q C_m \Phi_m \propto
\exp({-r^2/2a_\bot^2})F(u)$ with $F(u)=\Pi_{l=1}^{q}[u-b_\alpha]
$, the vortex function of $u$. Here $u$ refers to
$(x+iy)/{a_\bot}$ and $b_\alpha $ the position of
vortex~\cite{Ho}. A hole in the center of the gas was shown to be
generated with enhancing $\Omega$ in weak repulsive interaction
regime ~\cite{Fischer}, which was also notice in numerical
simulation~\cite{Kasamatsu}. Hence we assume that $\{b_\alpha\}$
form a regular lattice in which the vortex at the center site
carries $l$ units of angular momentum while the vortices at the
other sites merely carry a unit of angular momentum. Then $F(u)$
can be written as
\begin{equation}
F(u)=
u^{l-1}\prod_{\alpha=1}^q(u-b_\alpha),
\end{equation}
where $b_0=0$ is implied. The general form of $\Psi$ is likely
\begin{equation}
 \Psi=f(z)\Phi,  \quad
 \int{d^2r}\mid\Phi\mid^2=1,
\end{equation}
where $f(z)$ describes the density profile of the gas
along $z$-axis and  $\Phi=\bar{\Phi}/D(z)$ with
\begin{equation}
\bar{\Phi}=\exp{(-{r^2}/{2a_\bot^2}})u^{l-1}\prod_{l=1}^q(u-b_\alpha),
\end{equation}
and $D^2=\int{d^2r}\mid\bar{\Phi}\mid^2$, the normalization
constant for $\Phi$.
Now the number constraint
$\int{d^2r}\mid\Psi\mid^2$ becomes $\int{dz}\mid{f}\mid^2=N$.
Noting that in the complex coordinate
\begin{equation*}
L_z=\hbar(u\partial{u}-u^*\partial{u^*}),
\end{equation*}
we can obtain the expectation value of $L_z$,
\begin{equation}
\int\Psi^*L_z\Psi =
\hbar\int\bigl[(r/a_\bot)^2-1\bigr]\mid\Psi\mid^2.
\end{equation}
Then Eq.(\ref{eq:Kz}) becomes,
\begin{eqnarray}\label{eq:theKz}
K(z)&=&\Bigl[-\tilde{\mu}(z)+\frac{\hbar(\omega_{xy}-\Omega)\ave{r^2}_\Phi}{a_\bot^2}\Bigr]f^2
\nonumber
\\&&{}+\frac{M\omega_{xy}^2}{2a_\bot^2}\lambda{f^2}\ave{r^4}_\Phi+\frac{1}{2}gI_\Phi{f^4},
\end{eqnarray}
where $\tilde{\mu}(z)=\mu-\hbar\Omega-(M\omega_z^2z^2/2)$,
$\ave{r^2}_\Phi=\int{d^2r} r^2\mid\Phi\mid^2$,
$\ave{r^4}_\Phi=\int{d^2r} r^4\mid\Phi\mid^2$ and
$I_\Phi=\int{d^2r}\mid\Phi\mid^4$.
The minimization of the energy is performed by the variation of the parameter $\{b_\alpha\}$.

To evaluate $\ave{r^2}_\Phi$, $\ave{r^4}_\Phi$ and $I_\Phi$, we note that,
\begin{eqnarray}
\mid\bar{\Phi}\mid^2=e^{-H},\quad
H=H_1+H_2,
\end{eqnarray}
where $H_1$ $\equiv$
${r^2/a_\bot^2-2\sum_\alpha}{\ln\mid\vec{r}-\vec{b_\alpha}\mid}$
and $H_2\equiv{-2(l-1)\ln\mid\vec{r}\mid} $. By means of the Fourier
transform, one obtains~\cite{Ho},
\begin{eqnarray}
\nabla_\bot^2H_1&=&\frac{1}{\sigma^2}-\frac{4\pi}{v}\sum_{k\neq0}\cos\vec{k}\cdot\vec{r},\nonumber\\
\mid{\bar{\Phi}}\mid^2&=&r^{2(l-1)}e^{-(r^2/\sigma^2)}\prod_{k\neq0}\exp({\zeta_k{\cos\vec{k}.\vec{r}}}),\nonumber\\
\frac{1}{\sigma^2}&=&\frac{1}{a_\bot^2}-\frac{\pi}{v},\quad\quad\quad
{\zeta_k}=\frac{4\pi}{v\mid\vec{k}\mid}.
\end{eqnarray}
In the average-vortex approximation~\cite{Ho}, the
$|\bar{\Phi}|^2$ becomes
$|\bar{\Phi}|^2=r^{2(l-1)}\exp{-(r^2/\sigma^2)}=h(r)\exp{-(r^2/\sigma^2)}$;
here $h(r)\equiv{r^{2(l-1)}}$ is a slowly varying envelope
function, which gives rise to the components in higher Landau
Levels~\cite{Watanabe}; the $\Phi$ then becomes
$\mid{\bar{\Phi}}\mid^2=\bigl({r^{2(l-1)}
\exp{(-r^2/\sigma^2)}}\bigr)\big/\bigl({\pi\Gamma(l)\sigma^{2l}}\bigr)$.
Using the above result, we can easily evaluate the values.
$\ave{r^2}_\Phi=l\sigma^2$, $\ave{r^4}_\Phi=l(l+1)\sigma^4$, and
$I_\Phi={\xi}/({2\pi\sigma^2})$,
$\xi\equiv{(2l-1)!\big/\bigl(2^{2l-2}\bigl[(l-1)!\bigr]^2\bigr)}$.
Substituting these results into Eq(\ref{eq:theKz}), we have
\begin{eqnarray}\label{eq:3Kz}
&&K(z)=\Bigl[-\tilde{\mu}(z)+\frac{l\hbar(\omega_{xy}-\Omega)\sigma^2}{a_\bot^2}\Bigr]f^2
  \nonumber\\
&&\hspace{6mm}+\frac{\hbar\omega_{xy}\lambda}{2a_\bot^4}l(l+1)\sigma^4f^2+\frac{\hbar\omega_{xy}{a_{sc}}{a_\bot^2}}{\sigma^2}f^4\xi.
\end{eqnarray}
Minimizing the energy,  we obtain two equations,
\begin{eqnarray}
\frac{\partial{K(z)}}{\partial{f}}&=&2\Bigl[-\tilde{\mu}(z)+\frac{l\hbar(\omega_{xy}-\Omega)\sigma^2}{a_\bot}\Bigr]f
+4\frac{\hbar\omega_{xy}{a}_{sc}a_\bot^2
}{\sigma^2}\xi{f^3}\nonumber\\&&{}+\frac{\hbar\omega_{xy}\lambda}{a_\bot^4}l(l+1)\sigma^4f
=0,\nonumber\\
\frac{\partial{K(z)}}{\partial{\sigma}}&=&\frac{2l\hbar(\omega_{xy}-\Omega)\sigma}{a_\bot^2}f^2+\frac{2\hbar\omega_{xy}\lambda}{a_\bot^4}l(l+1)\sigma^3f^2\nonumber\\&&{}-\frac{2\hbar\omega_{xy}{a_{sc}a_\bot^2}}{\sigma^3}f^4\xi=0.
\end{eqnarray}
In the light that the above nonlinear equations are difficult to
solve analytically, we study those equations for the cases of
$\lambda=0$ and $\lambda\neq{0}$ so as to capture some physical
implications.
\begin{figure}[tbph]
\includegraphics[width=71mm]{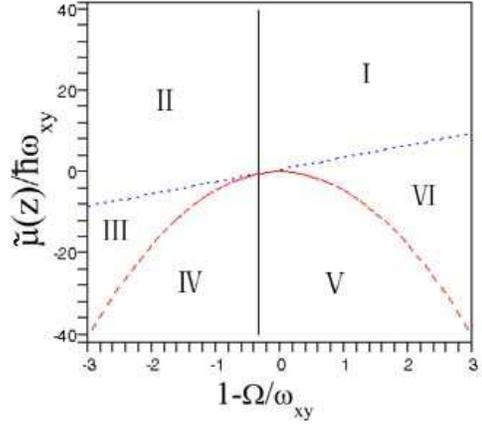}
\caption{\label{fig:phase}
(color on line) The phase diagram in the plane of
$\tilde{\mu}(z)$ versus $(1-\Omega/\omega_{xy})$
which is divided by the parabolic curve
$\tilde{\mu}(z)=-9\hbar\omega_{xy}{l}(1-\Omega/\omega_{xy})^2/10(l+1)\lambda$,
the oblique line
$\tilde{\mu}(z)=\hbar\omega_{xy}\bigl[2.5
l(l+1)\lambda+
3l(1-\Omega/\omega_{xy})\bigr]$
and the vertical line
$\tilde{\mu}(z)=-5\hbar\omega_{xy}(l+1)\lambda/3$.
}
\end{figure}

For $\lambda=0$, viz., the gas is confined in a harmonic trap,
\begin{eqnarray}\label{eq:sigma}
\Bigl(\frac{\sigma^2}{a_\bot^2}\Bigr)_0&=&\frac{\tilde{\mu}(z)^2}{3l\hbar(\omega_{xy}-\Omega)},
   \nonumber\\
\bigl(a_{sc}f^2\bigr)_0
 &=&\frac{\bigl[\tilde{\mu}(z)\bigr]^2}{9l^2\hbar^2\omega_{xy}(\omega_{xy}-\Omega)\xi}.
\end{eqnarray}
Substituting Eq.(\ref{eq:sigma}) into Eq.(\ref{eq:3Kz}),  we obtain,
\begin{equation}
\bigl[K(z)\bigr]_0=\frac{\bigl[\tilde{\mu}(z)\bigr]^3}{27\hbar^2\omega_{xy}(\omega_{xy}-\Omega)a_{sc}}\Bigl(-\frac{2}{l^2\xi}+\frac{1}{l^3\xi}\Bigr)\propto{\alpha},
\end{equation}
in which
$\alpha\equiv\bigl(-\frac{1}{l^2\xi}+\frac{1}{l^3\xi}\bigr)$
~\cite{Mueller}, for example,
$[\alpha]_{l=1}=-1$,
$[\alpha]_{l=2}=-{3}/{4}$, $[\alpha]_{l=3}=-{40}/{81}$ etc.
This implies that the energy of the system becomes larger with increase of $l$.
Thus the hole in the center of the system can not be formed in a harmonic trap,
which also be confirmed by variational wave function
$\Phi=Cr^m\exp({im\phi})\exp({\beta^2r^2/2a_\bot^2})$ by regarding
$\beta$ as variation parameter.
\begin{figure}[tbph]
\includegraphics[width=80mm]{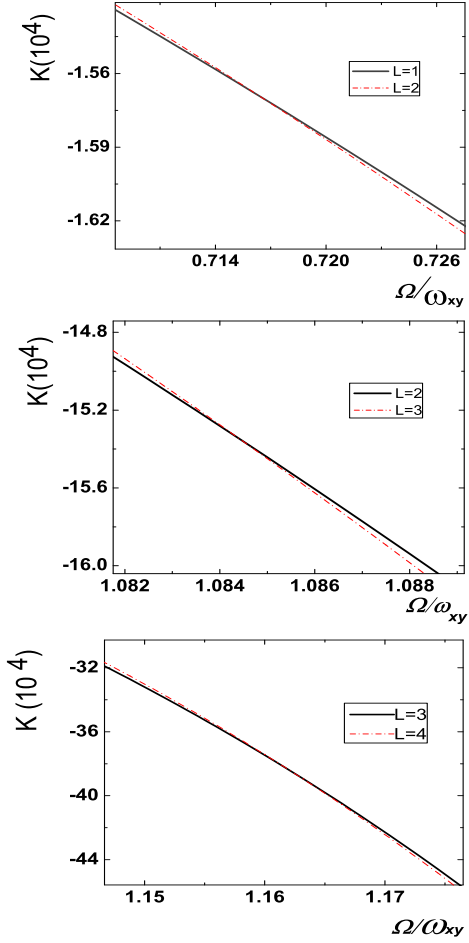}
\caption{\label{fig:K} (color on line) The curves of $K$ versus
$\Omega/\omega_{xy}$ in unit of
$(2\hbar\omega_{xy})^{3/2}M^{-1/2}\big/(2\omega_z a_{sc})$ for
different $l$ with parameters $N=3\times10^5$,
$\omega_{xy}=2\pi\times 65.6$, $\omega_z=2\pi\times11.0$,
$a_{sc}=53$\AA.}
\end{figure}

For $\lambda\neq0$, viz., the gas is confined in an anharmonic trap,
\begin{eqnarray}
\Bigl(\frac{\sigma^2}{a_\bot^2}\Bigr)_{\pm}&=&-\frac{3(1-\Omega/{\omega_{xy}})}{5(l+1)\lambda}\nonumber\\
&\pm&\frac{\sqrt{9(1-\Omega/{\omega_{xy}})^2
+\bigl[10(l+1)\lambda\tilde{\mu}(z)\bigr]\big/(l\hbar\omega_{xy})}}{5(l+1)\lambda},
\nonumber\\
 \xi{a_{sc}}f^2&=&
\frac{l\bigl(1-\Omega/{\omega_{xy}}\bigr)\sigma^4}{a_\bot^4}+\frac{\lambda{l}\bigl(l+1\bigr)\sigma^6}{a_\bot^6}.
\end{eqnarray}
Since $\sigma\geq a_\bot$, the condition for the existence of
solution of the above equations is plotted in Fig.\ref{fig:phase}.
In the regime I and II, only the root of $(\sigma^2/a_\bot^2)_+$
satisfies the condition $\sigma^2\geq{a_\bot^2}$. In the regime
III, both $(\sigma^2/a_\bot^2)_\pm$ satisfy that condition, which
implies the system has double stable states and two possible
configurations of vortex lattice can be formed. The vortices in
this regime become invisible for it is easy to vary between those
two configurations due to fluctuation. In the other regimes,
$(\sigma^2/a_\bot^2)_\pm$ do not satisfy that condition. In
Thomas-Fermi approximation, $\mu=M R_i^2\omega_i^2/2$, $R_i$ stand
for the maximum extents of condensate and $\omega_i$ for the
frequencies in three direction~\cite{Pethick}. According to the
experimental data~\cite{Hodby,Haljan}, the strait line
$\tilde{\mu}(z)=\mu-\hbar\Omega-M\omega_z^2z^2/2$ must cross the
regimes I, II and VI but not cross the other regimes. Thus the
condition  $\sigma^2\geq{a_\bot^2}$  becomes
$\tilde{\mu}(z)\geq\hbar\omega_{xy}\bigl[2.5l(l+1)\lambda+3l(1-\Omega/\omega_{xy})\bigr]$.
Note that for the same rotating frequency $\Omega$ with the
growing of $l$, the maximum extent of condensate along z-axis
$R_z$ will decrease,
\begin{equation}
R_z^2\propto\mu/\hbar\omega_{xy}-\Omega/\omega_{xy}-2.5l\bigl(l+1\bigr)\lambda-3l\bigl(1-\Omega/\omega_{xy}\bigr).
\end{equation}
For simplicity, we assume straight vortex lines~\cite{Fischer} so
that the (2D) density of vortex remains a constant along $z$-axis,
then we obtain  $K=2R_zK(0)$.

We plot the values of $K$ versus $\Omega/\omega_{xy}$ for
different $l$ in Fig.\ref{fig:K}. We can see that with
increasing  $\Omega$ the state becomes a mixed phase with a
multiply quantized vortex (hole) at the center of the cloud and
singly-quantized vortices around it~\cite{Jackson}. This assures
that a fast rotating Bose-Einstein condensate confined in a
quadratic-plus-quartic potential can generate a hole in the center
of the gas which is in agreement with the theoretical and
experimental studies~\cite{Kasamatsu,Bretin}. Our calculation
shows that if $\Omega$ increases slowly, an array of vortices is
formed at first (just like in harmonic trap); then a hole carrying
two unit of angular momentums is generated when
$\Omega_h/\omega_{xy}\approx0.716$, which keeps until it reaches
$\Omega/\omega_{xy}\approx1.084$. With the increasing of $\Omega$
the hole will absorb the singly-quantized vortices around it whose
angular momentum becomes larger and larger. Looking at the
positions of intersections in the three graphs in Fig.\ref{fig:K},
it is easy to find that the velocity of absorbing the
singly-quantized vortices becomes larger and larger, and finally a
giant vortex carrying all the  angular momentums will be formed.
When $\Omega/\omega_{xy}>1.084$, every configurations of the
vortex lattice remains for a short time, so it is difficult to
detect the vortex. Our result can also explain why no vortex lines
are detectable in the regime when $\Omega$ reaches to some large
value~\cite{Bretin}.

\paragraph{Conclusion:}
Employing the approach of Ref.~\cite{Ho},
we obtained the new form of the condensate wave function
which is essentially the Lowest Landau Level (LLL) wave function with a regular lattice of
vortices multiplied by a slowly varying envelope function.
Using the variational trial function, we obtained the critical value
$\Omega_h$ and proved that the fragility of the vortex lattice
could be enhanced by increasing $\Omega$.
This result is in agreement with the experiment result~\cite{Schweikhard}
where the gas was trapped in a harmonic potential.
In the experiment \cite{Schweikhard},
the vortex lattice remains ordered, but its elastic shear
strength is drastically reduced by increasing $\Omega$.
This implies that even a minor perturbation to the cloud can cause
the lattice to melt.
In the regime with very small $\lambda$, the quartic term also provide a perturbation
to destroy the vortex lattice, it is therefore invisible
when the $\Omega$ reaches some large value~\cite{Bretin}.

The work is supported by NSFC Grant No.10225419.

\end{document}